\documentclass[superscriptaddress,amsmath,amssymb]{revtex4}
\usepackage{graphicx} 

\begin{document}
  
\title{Entropy Production and the Pressure-Volume Curve of the Lung}

\author{Cl\'audio L. N. Oliveira}%
\email{lucas@fisica.ufc.br}%
\affiliation{Departamento de F\'isica, Universidade Federal do
  Cear\'a, 60451-970Research Topic Fortaleza, Cear\'a, Brazil}%

\author{Asc\^anio D. Ara\'ujo}%
\affiliation{Departamento de F\'isica, Universidade Federal do
  Cear\'a, 60451-970 Fortaleza, Cear\'a, Brazil}%

\author{Jason H. T. Bates}%
\affiliation{Department of Medicine, University of Vermont,
  Burlington, VT 05405}%
 
\author{Jos\'e S. Andrade Jr.}%
\affiliation{Departamento de F\'isica, Universidade Federal do
  Cear\'a, 60451-970 Fortaleza, Cear\'a, Brazil}%

\author{B\'ela Suki}%
\email{bsuki@bu.edu}%
\affiliation{Department of Biomedical Engineering, Boston University,
  Boston, MA 02215}%

\begin{abstract}
  We investigate analytically the production of entropy during a
  breathing cycle in healthy and diseased lungs. First, we calculate
  entropy production in healthy lungs by applying the laws of
  thermodynamics to the well-known transpulmonary pressure-volume
  ($P-V$) curves of the lung under the assumption that lung tissue
  behaves as an entropy spring-like rubber. The bulk modulus, $B$, of
  the lung is also derived from these calculations. Second, we extend
  this approach to elastic recoil disorders of the lung such as occur
  in pulmonary fibrosis and emphysema. These diseases are
  characterized by particular alterations in the $P-V$ relationship.
  For example, in fibrotic lungs $B$ increases monotonically with
  disease progression, while in emphysema the opposite occurs. These
  diseases can thus be mimicked simply by making appropriate
  adjustments to the parameters of the $P-V$ curve. Using Clausius's
  formalism, we show that entropy production, $\Delta S$, is related
  to the hysteresis area, $\Delta A$, enclosed by the $P-V$ curve
  during a breathing cycle, namely, $\Delta S=\Delta A/T$, where $T$
  is the body temperature. Although $\Delta A$ is highly dependent on
  the disease, such formula applies to healthy as well as diseased
  lungs, regardless of the disease stage.  Finally, we use {\it
    ansatzs} to predict analytically the entropy produced by the
  fibrotic and emphysematous lungs.
  \end{abstract}

\maketitle

\section{Introduction}
Although physics and biology developed as separate branches of
science, the role of physics in biology has assumed increasing
importance over the past century. This applies particularly to the
branch of physics known as thermodynamics. The laws of thermodynamics
are based on empirical evidence derived from the behavior of
macroscopic systems~\cite{Fermi1956}, and in this respect share
similarities with much of our knowledge about biological systems.
Indeed, in his seminal 1944 book ``What is life?'', Erwin
Schr\"odinger addressed the question of how living systems can
maintain order in apparent violation of the second law of
thermodynamics. He postulated that life is only possible if living
systems export entropy to their surroundings~\cite{Schrodinger1944}.
He even conjectured the existence of an ``aperiodic crystal''
containing the genetic information of living beings a decade earlier
than the discovery of DNA~\cite{Dyson1999}. His influential ideas
stimulated the development of molecular biology and many areas of
theoretical biology that are still being pursued today.

The field of thermodynamics has been greatly advanced by the advent of
the digital computer which provides the means to link thermodynamics
to microscopic mechanisms using the ideas of statistical mechanics in
situations that defy analytical calculation. This is also now finding
significant application in biology. For example, the microscopic
progress of fibrosis and emphysema in the lung has been linked to
pathologic changes in macroscopic lung function in terms of a
percolation process~\cite{Bates2007,Oliveira2014} and the fractal
dimension of nuclear chromatin has been found to provide a potential
molecular tool for cancer prognosis~\cite{Metze2013}. Additionally,
the connectivity of the brain has been studied in the framework of
complex networks~\cite{Reis2014} as well as the maximization of
entropy production~\cite{Seely2014}. These advances rely on extensive
numerical computation because of the highly nonlinear interactions
involved between the myriad components in these complex systems.

Regardless of these complexities, however, the laws of thermodynamics
must still hold. This applies in particular to the second law that
governs entropy. The very essence of a living system is continual
internal activity of a very ordered nature, but this activity
necessarily generates entropy which is the engine of disorder.
Nevertheless, living systems manage to maintain, throughout their
lifetimes, all electrical, chemical, and temperature gradients that
define their internal order~\cite{Annamalai2012}. Accordingly, living
systems must somehow export the entropy they generate to the
environment, as Schr\"odinger postulated~\cite{Schrodinger1944}. But
what happens if not all the entropy is exported? The remainder stays
within the system where its inescapable consequence must be a gradual
progression of the system toward malfunction (i.e., disease) and
eventual death. This raises two considerations that are paramount for
the life and health of an organism: 1) the rate at which entropy is
produced, and 2) the success with which that entropy is exported. In
this paper we focus on the first of these considerations in relation
to the lung, a well-defined thermodynamic system in the human body
that exchanges mass and energy continually with its surroundings.

The volume of fresh air inspired with every breath is a consequence of
the pressure generated by the respiratory muscles (principally the
diaphragm) and the elasticity of the lung tissues. The latter include
contributions from both the protein fibers of the extracellular matrix
and the surface tension of the air-liquid interface~\cite{Suki2011b}.
These events take place under essentially isothermal conditions
because temperature fluctuations deep in the lung are negligible even
though the temperature of the inspired air gradually increases from
ambient at the mouth to body temperature at some point along the
conducting airway tree~\cite{McFadden1985}. A thermodynamic model has
already been developed to predict the work done on the air-liquid
interface in the lung as a result of surface
tension~\cite{Prokop1999}, something that can change markedly in, for
example, acute respiratory distress syndrome~\cite{Gregory1991}. Our
focus here, however, is on pulmonary diseases that affect the elastic
protein fibers of the lung tissue, of which there are two main
examples.  {\it Pulmonary fibrosis} involves the excess production and
abnormal arrangement of protein fibers and thus causes the lung to
become stiffer than normal, while {\it emphysema} involves the
destruction of these fibers and so leads to a lung that is
correspondingly less stiff than normal~\cite{Levitzky1995}.
Currently, neither fibrosis nor emphysema can be cured, yet together
they constitute an enormous public health burden; fibrosis affects
approximately 5 million people worldwide~\cite{Meltzer2008}, while the
World Health Organization reports that emphysema led to the death of
more than 3 million people in 2012 alone~\cite{who}. Accordingly, in
the present study we propose a simple thermodynamic model of the
pressure-volume ($P-V$) relationship of the lung. We use this model to
calculate the entropy produced in the lung during normal breathing,
and then examine how this production is altered in pulmonary fibrosis
and emphysema.

\section{Thermodynamics of healthy lungs}
\begin{figure}[t]
  \begin{center}
    \includegraphics*[width=0.7\columnwidth]{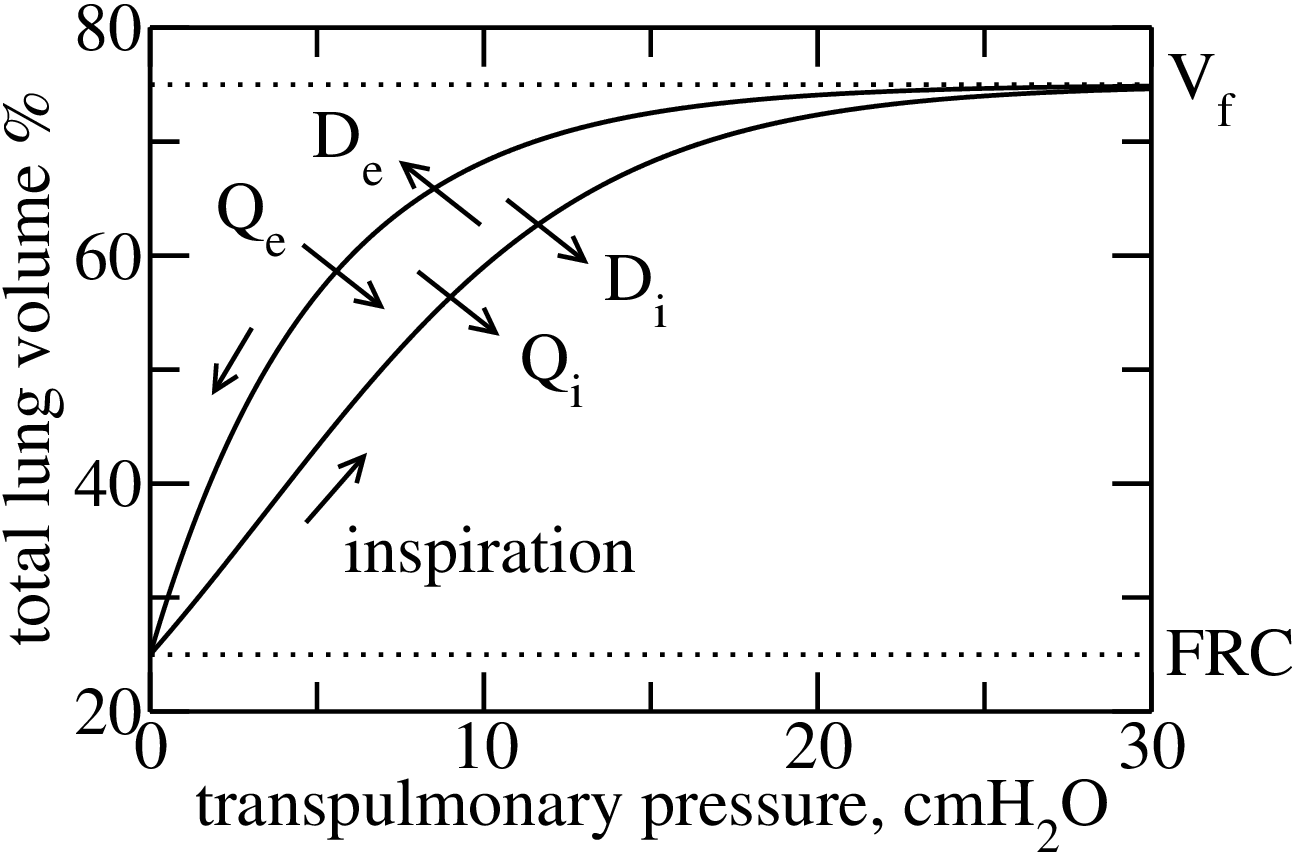}
    \caption{Typical transpulmonary pressure versus volume curves in
      healthy lungs. By considering the rubber approach, reversible
      heat $Q_i$ is released during inflation whereas it is absorbed
      during deflation $Q_e$. However, due to dissipation, heat is
      released during both inspiration and expiration, denoted by
      $D_i$ and $D_e$, respectively. The hysteresis is due to an
      asymmetry between the recruitment and derecruitment processes of
      collagen fibers, during inspiration and expiration,
      respectively.}\label{fig1}
  \end{center}
\end{figure}

We consider the lung as a purely elastic system with a state defined
by its volume ($V$). The equilibrium state is the value of $V$ at the
end of a relaxed expiration, known as Functional Residual Capacity
($FRC$), which is also taken here as the minimum $V$. During
inspiration, the respiratory muscles (principally the diaphragm)
create a pressure gradient across the lung, known as transpulmonary
pressure ($P$), that expands the lung to a volume $V_f$ that is
typically somewhat variable from breath to breath during normal
breathing but which has a maximum possible value known as Total Lung
Capacity ($TLC$). During expiration, $V$ is returned to $FRC$ by the
elastic recoil forces generated within the lung tissues during the
previous inspiration. Figure~\ref{fig1} shows typical $P$ versus $V$
($P-V$) curves for the lung. Such curves are well-known and can be
measured experimentally~\cite{Venegas1998}. (Note that $V$ here
represents the volume of air entering and leaving the lung during
breathing, not the volume of the lung tissue.)

The elastic recoil pressure of the lungs is generated as a result of
microscopic processes occurring within the lung tissue, such as the
stretching and unfolding of individual protein fibers. We assume here
that the lung tissue behaves similarly to rubber which is an elastic
material composed of long-chain polymers, called elastomers, that have
particular thermodynamic properties. For example, the Young's modulus
of rubber is proportional to absolute temperature, an intriguing
property that causes rubber to release heat when stretched as a result
of a corresponding decrease in entropy, and conversely to absorb heat
when returning toward equilibrium~\cite{Callen1985}.  Microscopically,
the decrease in entropy can be explained by progressively fewer
molecular conformations available for the elastomers as they stretch.
Conversely, the decreased entropy in the stretched state gives rubber
the ability to subsequently convert thermal energy into work as it
contracts against a load and its entropy increases. In this sense, a
rubber behaves somewhat like an ideal monatomic gas because neither
stores potential energy in the distortion of chemical bonds, but both
convert thermal energy into work on their
surroundings~\cite{Brown1963}.

The justification for considering lung elasticity to have an entropic
basis comes first of all from the fact that the principal structural
proteins in lung tissue are elastin and collagen, both of which are
organized into long tortuous fibers. For both
elastin~\cite{Baldock2011} and collagen~\cite{Buehler2007} these
fibers have been modeled, for modest stretches, as worm-like chains
that behave like entropic springs (although at high levels of strain
both fiber types begin to store elastic energy in their molecular
bonds). Collagen is at least 100 times stiffer than elastin, so for
simplicity we will assume that collagen fibers are actually infinitely
stiff so that worm-like chain entropy applies only to the elastic
fibers. Entropy also applies to collagen fibers but for a different
reason, as follows. The collagen and elastic fibers form an
essentially random network in which the stress at low strain is borne
almost exclusively by the elastin fibers and the collagen fibers are
flaccid and wavy. As strain ($V$) increases the collagen fibers become
taught and thus prevent those elastin fibers in their immediately
vicinity from being able to stretch. This gives rise to a progressive
stiffening of the entire tissue as $V$ increases, as seen in the $P-V$
curve (Fig.~\ref{fig1}) and as modeled previously in
1D~\cite{Maksym1997}. However, the collagen fibers in 3D lung tissue
are not entirely constrained in their orientations but rather may
assume different directions as a result of thermal
motion~\cite{Bates1998}. At equilibrium these fiber directions may be
quite random but as the tissue stretches the fibers become oriented
preferentially in the direction of local strain. This reduces the
number of possible configurations of the fibers within the tissue
matrix and hence reduces their entropy.  Assuming that the fibers
resist being oriented in the direction of strain to a degree that is
proportional to absolute temperature, $T$, collagen recruitment can
also be modeled as an entropic process similar to the stretching of
rubber.

We can thus reason that the collagen and elastin fibers in lung tissue
ought to behave together as an entropically elastic material. Note,
however, that these fibers do not undergo their thermodynamic
excursions within the living lung in isolation but rather exist under
essentially isothermal conditions because the metabolic processes of
life, and especially the heat-exchanging capacity of the circulating
blood, maintain core body temperature at an even 37$^{\circ}$ C.
Consequently, these fibers have the capacity to exchange heat with
their environment and thus to dissipate energy, which occurs as a
consequence of the frictional heat that is generated as the fibers are
continually jostled by thermal motion. Thus, an amount of heat energy
$D_i$ is released irreversibly to the surroundings as a result of
frictional losses during inspiration. Similarly, during expiration an
amount $D_e$ is released irreversibly as frictional losses. Note that
these frictional heats are different to the heats released during
inspiration and imported during expiration as a result of entropic
changes, namely $Q_i$ and $Q_e$, respectively. In other words, even
though the macro-configurations of the collagen and elastin fiber
systems may be identical at the end of each expiration, their
micro-configurations are different from breath-to-breath, and
frictional energy is dissipated in moving from one end-expiratory
micro-configuration to the next.

Clausius formulated the Second Law as follows:
\begin{equation}
  N = S - S_0 - \int\frac{dQ}{T},
  \label{Eq:Clausius}
\end{equation}
where $N>0$ is the so-called uncompensated transformation, which is
the entropy due to irreversible processes within the system. $S$ and
$S_0$ are the entropies of the final and initial states and $T$ is the
absolute temperature. The last term identifies any exchange of heat
with the environment. Hence, $\Delta S_i=S-S_0$ represents the entropy
production during an irreversible process that moves the system from
the initial to the final state. In our case, since the lung returns at
the end of each breath to the same volume, $FRC$, at the same
temperature, $T$, the entropy of the tissues at the end of a breath
cycle should be the same as at the end of the previous cycle. This
implies that the entropy produced by the irreversible processes is
exported to the environment, principally the heat bath provided by the
circulation.

Now, the change in entropy $\Delta S_r$ around the cycle due to the
any alterations in the configurations of the elastin and collagen
fibers must be zero because we consider the elastic properties of lung
tissue to be conservative. In other words, the last term in
Eq.~\ref{Eq:Clausius} cancels during over cycle:
\begin{equation}\nonumber
  \Delta S_r = \frac{-Q_i+Q_e}{T} = 0,
\end{equation}
which also means that the change in entropy of the system is entirely
due to the frictional work, $N=\Delta S_i$, which is given by
\begin{equation}
  \Delta S_i = \frac{D_i+D_e}{T} > 0,
  \label{Eq:dSi1}
\end{equation}
where $D_i$ and $D_e$ are the amounts of frictional energy dissipated
during inspiration and expiration, respectively.

On the other hand, the sum of $D_i$ and $D_e$ is the total frictional
energy dissipated around the breath cycle, which equals the hysteresis
area of the $P-V$ loop (Fig.~\ref{fig1}). This area is
\begin{equation}
  \Delta A = \int_{FRC}^{V_f} P_idV + \int_{V_f}^{FRC} P_edV = D_i+D_e,
  \label{Eq:dA}
\end{equation}
where $P_i$ and $P_e$ are simply $P$ during inspiration and
expiration, respectively. Substituting into Eq.~\ref{Eq:dSi1} then
gives
\begin{equation}
  \Delta S_i = \frac{A}{T},
  \label{Eq:dSi2}
\end{equation}
where $\Delta S_i$ is positive since $\Delta A$ is positive. Notice
that if the area between the curves vanishes, $N$ in
Eq.~\ref{Eq:Clausius} also vanishes as predicted by the Clausius
formulation for reversible processes. Equation~\ref{Eq:dSi2} shows
that the energy dissipated during each breathing cycle can be linked
directly to entropy production, $\Delta S_i$, which is exported to the
environment with each breath.

\begin{figure}[t]
  \begin{center}
    \includegraphics*[width=0.6\columnwidth]{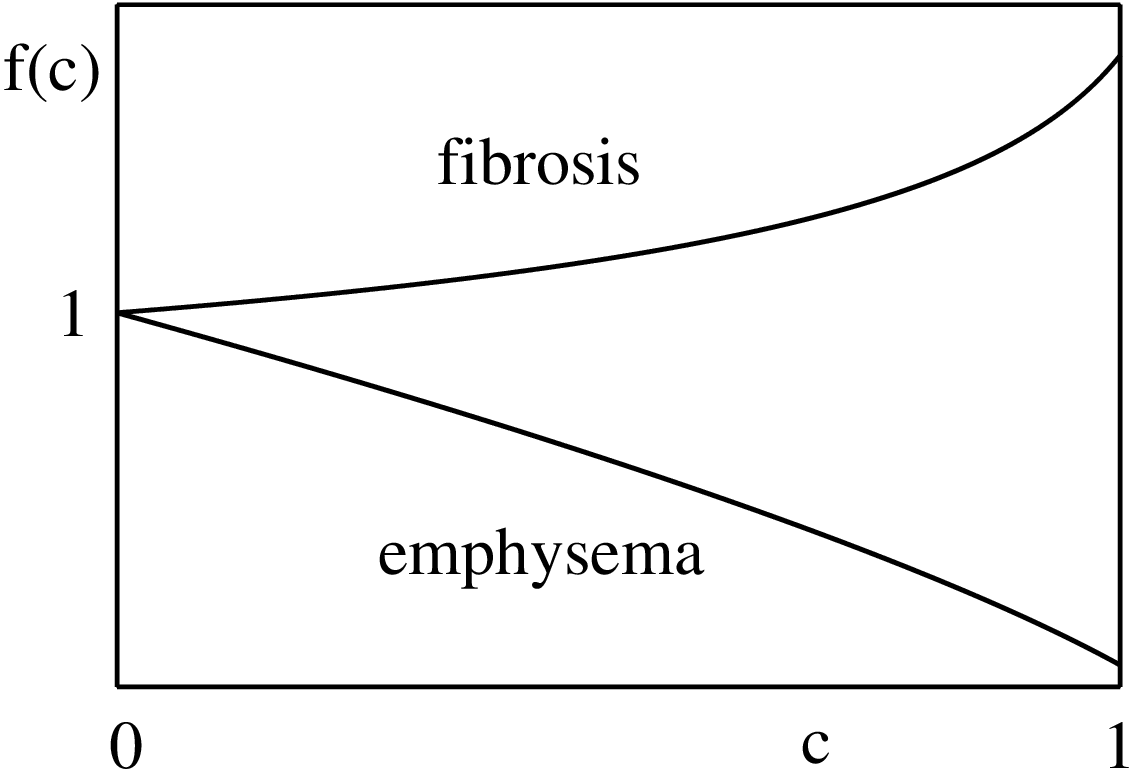}
    \caption{Function $f(c)$ behaves differently for different
      diseases. In fibrosis, it increases with fraction of lung
      parenchymal tissue affected by disease, $c$, while, in
      emphysema, it decreases with the disease stage. Although, the
      explicit nature of $f(c)$ is unknown, it should start from unity
      and change monotonically with $c$.}\label{fig2}
  \end{center}
\end{figure}

\section{Analytical fittings of the transpulmonary $P-V$ curves}
We can obtain a formula for $\Delta S_i$ from analytical expressions
for the inspiratory and expiratory $P-V$ curves shown in
Fig.~\ref{fig1}. These curves can be fitted with sigmoidal and
exponential functions, respectively, as follows~\cite{Venegas1998},
\begin{eqnarray}\nonumber
  V_i &=& \frac{V_f}{1+e^{-\frac{P-a}{b}}},\\
  V_e &=& V_f - \Delta Ve^{-P/k},
\label{Eq:Vp}
\end{eqnarray}
where $V_i$ and $V_e$ represent $V$ during inspiration and expiration,
respectively. The difference $\Delta V=Vf-FRC$ is the change in lung
volume during a breath, and is usually referred to as tidal volume.
Note that $V_f$ is substantially smaller than $TLC$ during normal
resting breathing. During inspiration, $V_i$ begins at its minimum
value of $FRC$ (when $P=0$) and increases to $V_f$, in which case the
parameter $a=b\ln(\Delta V/FRC)$ represents the inflection point of
the sigmoid. The parameter $b$ governs the slope of the sigmoid at its
inflection point; the larger is $b$ the smaller is the slope. The
exponential equation for $V_e$ in Eqs.~\ref{Eq:Vp} is governed by a
single parameter, the exponent $k$ that, like $b$, governs the rate of
change of volume with pressure except this time during expiration.

Rewriting Eqs.~\ref{Eq:Vp} explicitly in terms of $P$ gives
\begin{eqnarray}\nonumber
  P_i &=& a + b\ln \left(\frac{V}{V_f - V}\right),\\ 
  P_e &=& k\ln \left(\frac{\Delta V}{V_f - V}\right).
\label{Eq:Pv}
\end{eqnarray}
Finally, integrating these equations with respect to $V$ and
substituting into Eq.~\ref{Eq:dA} gives
\begin{equation}
  \Delta S_i = \frac{1}{T}\left[bV_f\ln\left(\frac{V_f}{FRC}\right)
    -k\Delta V\right].
  \label{Eq:dSi3}
\end{equation}
This equation defines the entropy produced (and exported) by the lung
tissue during a single breathing cycle as a function of the tidal
volume, $V_f$. The parameters $b$, $k$, $FRC$ and $T$ can be taken to
be constants for a normal adult lung, but may vary with disease.
Remarkably, the first term on the left-hand side of Eq.~\ref{Eq:dSi3}
is homomorphic to the change in entropy of an ideal gas when its
volume increases from $V_A$ to $V_B$ under isothermal conditions,
namely, $\Delta S_{\mathrm{ideal\,gas}}=(P_BV_B/T)\ln(V_B/V_A)$.

\section{Bulk modulus}
The bulk modulus of the lung is the inverse of its specific compliance
and characterizes its elastic properties; the larger the bulk modulus,
the stiffer (less compliant) the lung. The bulk modulus $B$ is thus
defined as~\cite{Parameswaran2011}
\begin{equation}\nonumber
  B=V\frac{dP}{dV}.
\end{equation}

Using Eqs.~\ref{Eq:Pv} one finds that $B$ during inspiration and
expiration is given by,
\begin{eqnarray}\nonumber
  B_i &=& b\frac{V_f}{V_f-V},\mathrm{\,\,and}\\ \nonumber
  B_e &=& k\frac{V}{V_f-V},
\end{eqnarray}
respectively. Because of the nonlinear $P-V$ relationships, $B$
changes with $V$ during both inspiration and expiration. For
simplicity, therefore, we will consider a representative $B$ at the
halfway point of the breath, i.e., at $V=V_f/2$, which gives $B_i=2b$
and $B_e=k$. Moreover, it is always observed experimentally that
$B_i>B_e$, so in the following we will use $k=b$, which satisfies this
condition.

\section{Applying the model to fibrotic and emphysematous lungs}
It has been observed that Eqs.~\ref{Eq:Pv} also provide good fits to
the P-V curve of both fibrotic~\cite{Ferreira2011} and
emphysematous~\cite{Soutiere2004,Rial2014} lungs. The altered $P-V$
curves in these diseases can thus be mimicked simply by adjusting the
parameters in Eqs.~\ref{Eq:Pv}. In fibrosis the lung becomes stiffer
so patients need to apply more pressure to inspire a smaller volume of
air. In emphysema the loss of lung elasticity increases $FRC$ due to
the outward recoil of the chest wall. We therefore model fibrosis by
increasing $b$, while emphysema is modeled by decreasing $b$.
Specifically, we let $b$ vary with disease state according to
\begin{equation}\nonumber
  b = k = b_0f(c),
\end{equation}
where $c$ is the fraction of lung parenchymal tissue affected by
disease (a measure of disease severity) and $f(c)$ is a function that
starts from 1, at $c=0$, and increases (decreases) monotonically with
$c$ for fibrosis (emphysema), and $b_0$ is the value of $b$ for a
healthy lung.  Thus, $b$ starts at $b_0$ and changes monotonically
either up or down as the disease progresses. Figure~\ref{fig2} shows
schematically how $f(c)$ changes for fibrosis and emphysema.

Another important physiological change that occurs in both fibrosis
and emphysema is that $V_f$ also changes with disease progression, so
$V_f$ is also a function of $c$. Specifically, $V_f(c)$ decreases in
fibrosis and increases in emphysema. This has the effect of
essentially creating a smaller or larger lung, respectively, which
means that the ratio of $V_f$ to $FRC$ in Eq.~\ref{Eq:dSi3} remains
unchanged.  $V_f$ and $FRC$ thus change in the same proportion
according to the function $g(c)$ thus:
\begin{eqnarray}\nonumber
  FRC(c) &=& FRC_0g(c),\mathrm{\,\,and}\\ \nonumber
  V_f(c) &=& V_{f0}g(c).
\end{eqnarray}
Where $FRC_0$ and $V_{f0}$ are the healthy values for $FRC$ and $V_f$,
respectively. Like $f(c)$, $g(c)$ also starts from 1, at $c=0$, and
changes monotonically with $c$ but in the reverse direction. That is,
$f(c)$ increases in fibrosis while $g(c)$ decreases to account for the
fractional change in lung volume that occurs with disease progression.
Conversely, $g(c)$ decreases in emphysema while $g(c)$ increases.

We are now in a position to describe how the entropy production per
breathing cycle changes as disease evolves. Consider, for example, the
case of a deep inspiration to $TLC$ (i.e., $V_f=TLC_0$). We can then
compare the behavior of $P-V$ curves in diseased lungs to healthy
lungs. This gives, from Eq.~\ref{Eq:dSi3},
\begin{equation}
  \Delta S_i(c) = \frac{b_0f(c)g(c)}{T} 
  \left[TLC_0\ln \left( \frac{TLC_0}{FRC_0} \right) - \Delta V_0 \right],
  \label{Eq:dSi4}
\end{equation}
where $\Delta V_0=TLC_0-FRC_0$.

Equation~\ref{Eq:dSi4} shows that the entropy produced in the lungs
over the breath cycle changes with disease by an amount given simply
by multiplying $\Delta S_i$ (from Eq.~\ref{Eq:dSi3}) by the product of
$f(c)$ and $g(c)$. This shows how the alteration of bulk modulus, as
well as the alteration in the parameter $b$ in disease plays a role in
entropy production. Additionally, one interpretation of entropy
production is that its increase in a given disease condition signifies
a less efficient mechanical function for the lung and more of the
elastic recoil is converted into heat.

\begin{figure}[t]
  \begin{center}
    \includegraphics*[width=0.7\columnwidth]{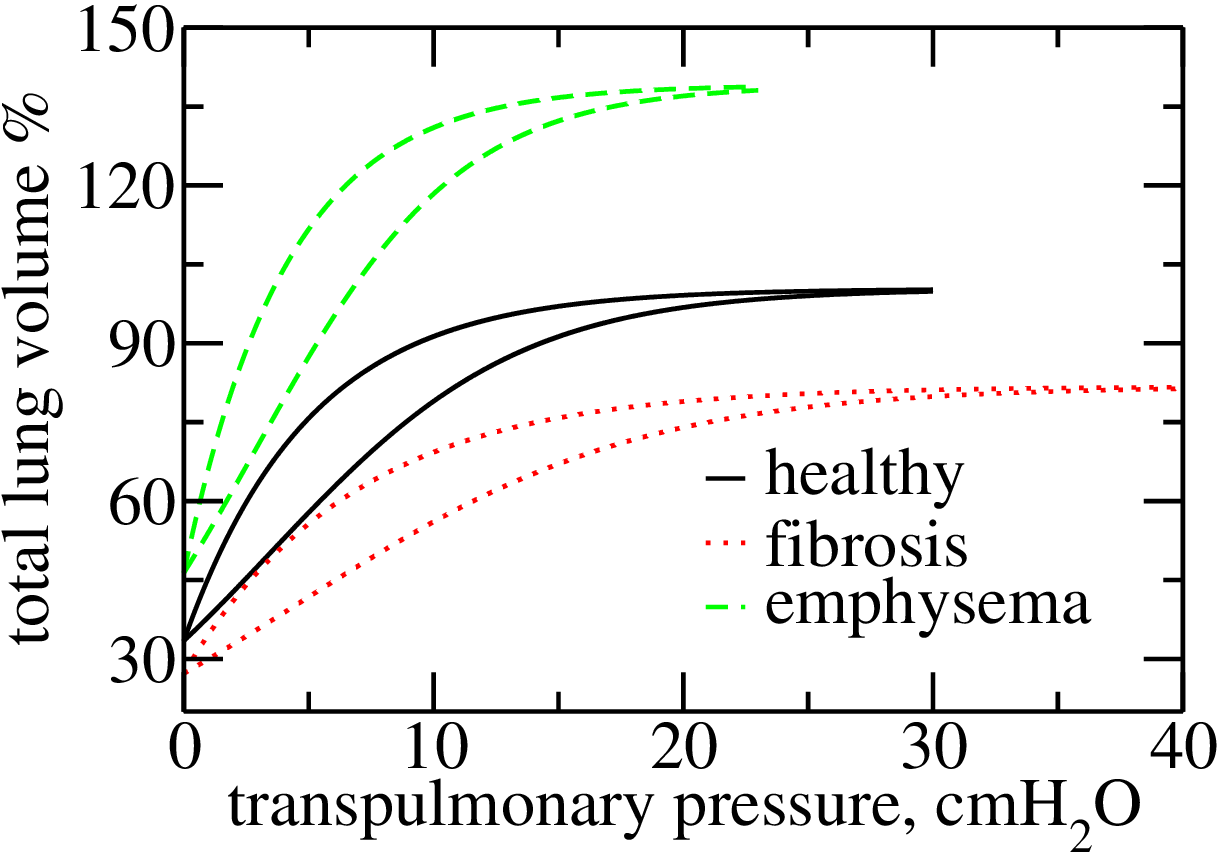}
    \caption{Transpulmonary $P-V$ curves for healthy, fibrotic, and
      emphysematous lungs, for $c=0.3$. For fibrosis, the curves are
      obtained for $f_{fi}(c)=(1-c)^{-\alpha}$, where $\alpha=0.6$,
      and $g_{fi}(c)=(1-c)^{\gamma}$, where $\gamma=0.4$. For
      emphysema, $f_{em}(c)=e^{-\left(\frac{c}{\beta}\right)^2}$,
      where $\beta=0.9$, and
      $g_{em}(c)=e^{\left(\frac{c}{\kappa}\right)^2}$, with
      $\kappa=0.7$. For these curves we use the following parameters,
      $b_0=5.0\mathrm{\,cmH2O}$, $FRC_0=2.5\mathrm{\,liters}$, and
      $TLC_0=6.5\mathrm{\,liters}$.}\label{fig3}
  \end{center}
\end{figure}

\section{Ansatz for $f(c)$ and $g(c)$}
It remains to define $f(c)$ and $g(c)$ for either fibrosis or
emphysema.  Conceivably, these functions could be determined by
analyzing $P-V$ curves at different stages of the disease, but this
has yet to be done. Alternatively, the functions could be guessed at
on the basis of the behavior of a computational model of disease
progression, such as the percolation model we have previously
investigated~\cite{Oliveira2014}. To keep things simple at this point,
however, we take here a simple empirical approach by first noting that
$f(c)$ and $g(c)$ should start at unity and change monotonically with
the progression of disease. Furthermore, it is known that the symptoms
of fibrosis only become apparent when about 30\% of the lung is
affected, while emphysema symptoms may be noticed at an earlier stage.
In other words, the function $f(c)$ for fibrosis should not change
much until $c\approx 0.3$, whereas in emphysema symptoms may occur for
$c\gtrsim 0$.

Accordingly, we make the following assumptions for the $f$ functions
for fibrosis, $f_{fi}(c)$, and for emphysema, $f_{em}(c)$:
\begin{eqnarray}\nonumber
  f_{fi}(c) &=& \left(1-c\right)^{-\alpha},\mathrm{\,\,and}\\ \nonumber
  f_{em}(c) &=& e^{-\left(\frac{c}{\beta}\right)^2},
\end{eqnarray}
The equation for $f_{fi}(c)$ mimics the fact that fibrosis progresses
slowly at early stages but grows faster as the affected tissue nears
the percolation threshold in the lung. The equation for $f_{fi}(c)$
and $f_{em}(c)$ captures the behavior of the bulk modulus of
emphysematous lungs as found in previous
studies~\cite{Oliveira2014,Parameswaran2011}.

The function $g(c)$, which defines how $TLC$ and $FRC$ change with the
disease, is actually harder to predict without experimental data. It
has been reported, however, that the $P-V$ area or the dissipation
during breathing increases both in fibrosis~\cite{Manali2011} and
emphysema~\cite{Ito2004}. Here, for simplicity, we apply a similar
analytical approach as that used for $f(c)$. That is,
\begin{eqnarray}\nonumber
  g_{fi}(c) &=& \left(1-c\right)^{\gamma},\mathrm{\,\,and}\\ \nonumber
  g_{em}(c) &=& e^{\left(\frac{c}{\kappa}\right)^2},
\end{eqnarray}
for fibrosis and emphysema, respectively. If $\gamma$ and $\kappa$ are
positive, then $g_{fi}(c)$ decreases while $g_{em}(c)$ increases with
$c$. Besides, in order for $\Delta A$ to increase with $c$ for both
diseases as reported in the literature~\cite{Manali2011,Ito2004}, the
following conditions must be met: $\alpha>\gamma$ and $\beta>\kappa$.

\begin{figure}[t]
  \begin{center}
    \includegraphics*[width=0.45\columnwidth]{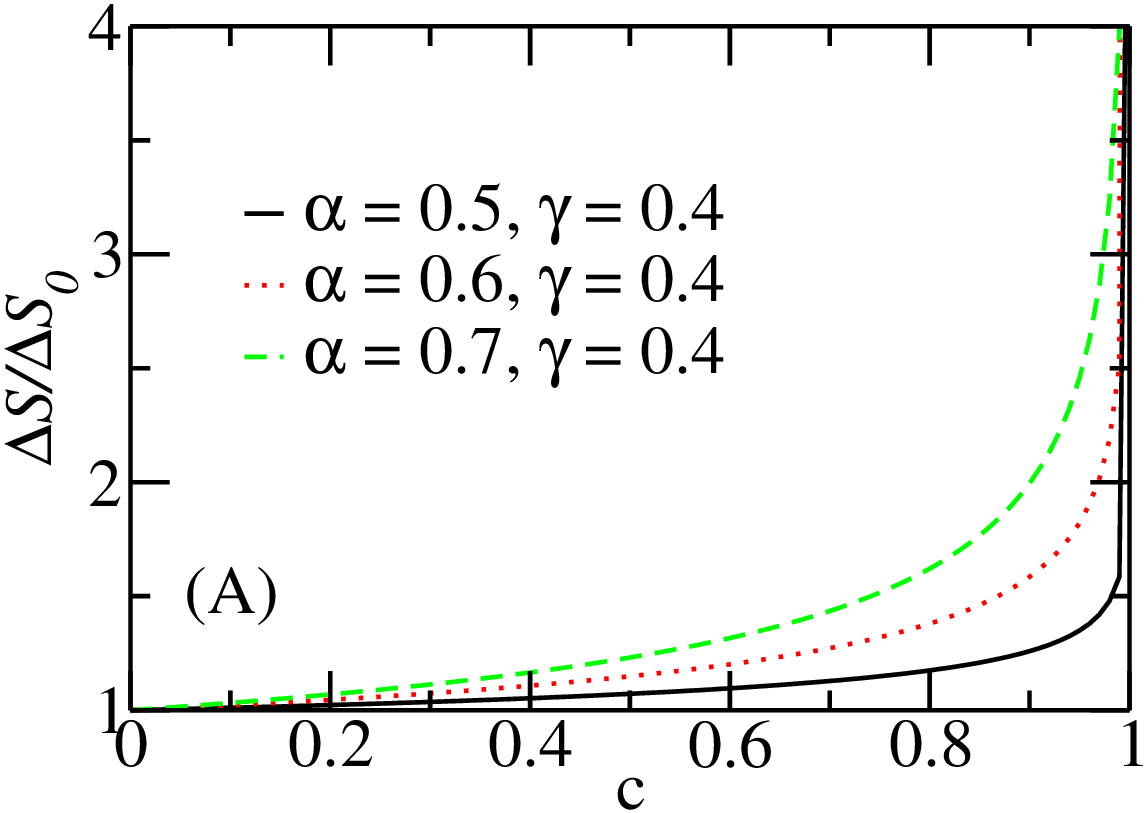}
    \includegraphics*[width=0.45\columnwidth]{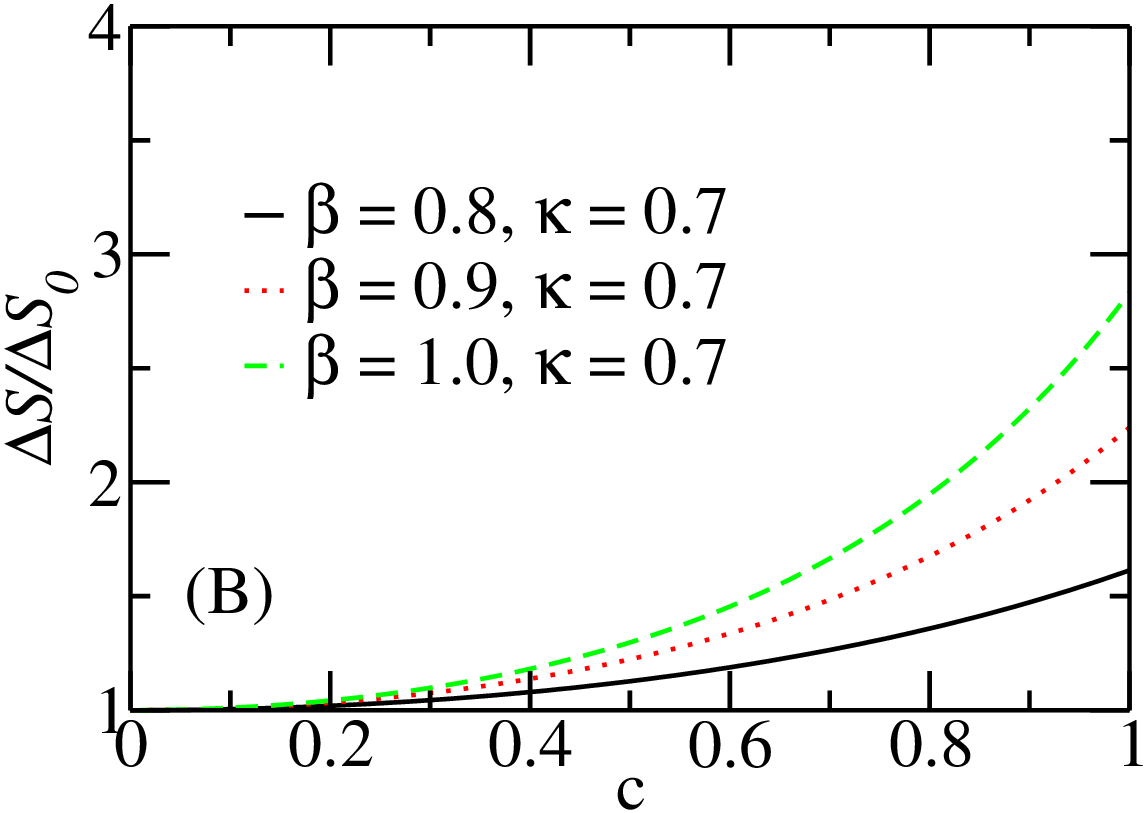}
    \caption{The entropy production per breathing cycle as a function
      of $c$. For different values of parameters, $\alpha$, $\beta$,
      $\gamma$, and $\kappa$, in fibrosis (A) and emphysema (B).
      Notice that the entropy production is normalized with its value
      at $c=0$. The other parameters are the same as those used in
      Fig.~\ref{fig3}.}\label{fig4}
  \end{center}
\end{figure}

Figure~\ref{fig3} shows the $P-V$ curves, plotted using these
analytical expressions, for healthy lungs as well as fibrotic and
emphysematous lungs, for several sets of parameters.
Figure~\ref{fig4} shows the entropy production as a function of $c$
for fibrosis (Fig.~\ref{fig4}(A)) and emphysema (Fig.~\ref{fig4}(B)).
Notice the sudden increase of entropy production for $c>0.8$ in
fibrosis, which suggests that in end-stage disease respiration becomes
highly inefficient as much of the elastic energy stored in the fibers
by the respiratory muscles is dissipated as heat. On the other hand,
in emphysema, the entropy production increases much slower, suggesting
a more gradual deterioration of the efficiency of the lung.

\section{Limitations}
The model developed here has several important limitations. First, we
neglect the contribution of surface tension at the air-liquid
interface to the mechanical behavior of the lung. However, surface
tension and, more importantly, airway closure and re-opening are
important issues at low lung volumes and in diseases that are
accompanied by edema formation. The effect of surface tension is much
less in the normal lung and in emphysema and fibrosis than in acute
lung injury. We also neglect the energetic contribution of collagen to
lung elastic recoil. Instead, we argue that fiber alignment and
recruitment can be modeled as a change in configuration, an assumption
still that has to be experimentally verified. We also neglect the
explicit mechanisms at the microscale that likely contribute to
entropy production in the tissue. In several previous studies, it has
been argued that polymer reptation~\cite{Suki1994}, fiber
alignment~\cite{Bates1998}, fiber-fiber
interactions~\cite{Mijailovich1993} as well as collagen-proteoglycan
interactions~\cite{Suki2011a} might contribute to the dissipative
processes in the lung tissue.

\section{Conclusions}
We have developed a thermodynamic model of the mechanics of breathing
that gives a central role to entropic changes in the lung tissue. We
used this model to predict how fibrosis
and emphysema alter entropy production in the lung over the
breathing cycle. Interestingly, our results show that both fibrotic
and emphysemathous lungs produce more entropy than healthy lungs. The
sicker is the lung, the more entropy is produced. This is a
consequence of the hysteresis area, enclosed by the $P-V$ curves,
which is increased in both diseases. The question remains as to
whether all the entropy that is produced in this manner is actually
exported to the environment, or part of it is retained in the
lung so that, over time, the organized structure of the lung
deteriorates as a manifestation of aging and/or disease.

\section{Acknowledgments}
This work was supported by CNPq, CAPES, FUNCAP and NIH HL124052.

\end{document}